\documentclass[showkeys,amsmath,amssymb,preprint,nofootinbib]{revtex4}
\usepackage{graphicx}
\usepackage{epsfig}
\usepackage{bm}
\usepackage{float}
\usepackage[T1]{fontenc}
\usepackage[utf8]{inputenc}
\usepackage{orcidlink}
\usepackage{graphicx}
\usepackage{bm}
\usepackage{amsmath}
\usepackage{xcolor}
\DeclareUnicodeCharacter{2212}{\textendash}
\def\beq{\begin{eqnarray}}

\def\eeq{\end{eqnarray}}

\begin{document}
\title{A new approach to $P-V$ phase transitions: Einstein gravity and holographic type dark energy}

\author{Miguel Cruz\orcidlink{0000-0003-3826-1321}$^{a,}$\footnote{miguelcruz02@uv.mx}, Samuel Lepe\orcidlink{0000-0002-3464-8337}$^{b,}$\footnote{samuel.lepe@pucv.cl} and Joel Saavedra\orcidlink{0000-0002-1430-3008}$^{b,}$\footnote{joel.saavedra@pucv.cl}}

\affiliation{$^{a}$Facultad de F\'{\i}sica, Universidad Veracruzana 91097, Xalapa, Veracruz, M\'exico \\
$^{b}$Instituto de F\'\i sica, Pontificia Universidad Cat\'olica de Valpara\'\i so, Casilla 4950, Valpara\'\i so, Chile}

\date{\today}

\begin{abstract}
In the framework of Einstein's gravity, we study the thermodynamic equation state, $P=P(V,T)$, associated with a flat Friedmann-Lemaitre-Robertson-Walker (FLRW) universe. In this scenario, we consider the components of the dark sector as non-interacting fluids that dominate the universe's energy content at late times. Under these circumstances, the functional structure of the cosmological coincidence parameter plays a relevant role in admitting first-order $P-V$ phase transitions; specifically, the dark energy density and the coincidence parameter must be given in terms of the radius of the apparent horizon.
\end{abstract}

\keywords{dark sector, FLRW cosmology, apparent horizon, Einstein gravity}

\maketitle

\section{Introduction}
\label{sec:intro}
Black holes have been a very important environment for understanding the linkage between general relativity and thermodynamics for several decades. This has led to important results such as relating macroscopic properties associated with thermodynamic systems with the geometry of black holes, especially through the event horizon \cite{bekenstein, haw, haw2, dolan, haykoda}. However, this intimate connection seems to emerge from Einstein's equations since they can be interpreted as the first law of thermodynamics \cite{jacobson}. Consequently, it is natural to think that the results obtained from the black hole dynamics can be translated into the cosmological scenario by replacing the usual fixed horizon with a dynamical object; in this latter construction, such horizon corresponds to the so-called apparent horizon \cite{helo0}. This provides a natural framework to realize the universe as a thermodynamical system, see for instance references given in \cite{bekenstein2} as a first approach to this idea. It is worth mentioning that the nature of the cosmological horizon dictates the behavior of the thermodynamic quantities associated with it, i.e., there are subtle differences between the thermodynamic of a contracting cosmology and an expanding one, we refer the reader to Refs. \cite{helou, posit}, where these important distinctions between cosmologies were established, the difference in the thermodynamics emerging from black holes and white holes was also studied. From now on, we restrict ourselves to the case in which the cosmological principle prevails, and the universe's matter content is trapped by the apparent horizon surface in an expanding evolution.

Based on the above discussion, Einstein's theory represents a robust scenario to explore the thermodynamic nature of the observable universe. Contrary to what is generally found in the literature under the thermodynamic system interpretation for an FLRW universe, see for instance \cite{modified} see also \cite{various}, where modified theories of gravity are considered; in this work, we show that phase transitions in Einstein gravity are allowed by only assuming more than one fluid as matter sector. We consider the dark sector components as the universe's dominant content and relate them using the cosmological coincidence parameter; this is the first and main consideration of our formulation. The dark energy and dark matter fluids are necessary to induce changes in the usual equation of the state of the FLRW universe, namely 
\begin{equation}
    P = \frac{T_{A}}{2R_{A}}+\frac{1}{8\pi R_{A}^{2}}.\label{eq:usual}
\end{equation}
However, adopting the standard thermodynamic description \cite{callen}, we find that the existence of critical points (and phase transitions) is determined by some conditions that are sensitive only under modifications that can be expressed in terms of the geometry of the apparent horizon, specifically its radius, $R_{A}$, as we will see later. At first glance, this closure condition enforces dark energy to be of the holographic type to write also the cosmological coincidence parameter in terms of $R_{A}$. Despite the limited this may seem, we have at hand several cases of interest that lead to phase transitions since we also provide some dark energy models physically motivated beyond the holographic scheme that fulfills the condition mentioned before. The approach presented here does not intend to oppose the idea of the necessity of modified gravity scenarios to study the thermodynamics of the universe but to show that Einstein's theory is capable of exhibiting a rich structure from the thermodynamic point of view with a simple refinement in the description of the matter sector.

The organization of the work is given as follows: in Section \ref{thermo}, we describe the apparent horizon of an FLRW universe. We establish relevant quantities associated with its thermodynamic description, such as surface gravity, temperature, and the unified first law. This is useful for relating the work density to the thermodynamic pressure, $P$. In Section \ref{pv}, we specialize the apparent horizon results to the case $k=0$ and use the standard procedure to establish the existence of $P-V$ phase transitions from the equation state of the universe, $P(V,T)$. We apply the method to the case where the Friedmann equations describe two non-interacting fluids only related by the cosmological coincidence parameter. We describe the nature of the phase transition by exploring the behavior of the Gibbs free energy. The final comments of the work are given in Section \ref{conclusion_f}. $G=c=1$ units will be used throughout the work.

\section{FLRW universe and apparent horizon}\label{thermo}
This section will briefly revise the thermodynamic description associated with the apparent horizon in an FLRW universe \cite{helo0}; such description is borrowed from the black hole dynamics \cite{haykoda}. In our formulation, we take into account two non-interacting cosmological fluids. The FLRW line element is usually given as
\begin{equation}
    ds^{2} = -dt^{2}+a^{2}(t)\left[\frac{dr^{2}}{1-kr^{2}}+r^{2}(d\theta^{2}+\sin^{2}\theta d\varphi^{2}) \right], \label{eq:flrw}
\end{equation} 
and it is well known that such metric can be written equivalently as follows using a warped product
\begin{equation}
    ds^{2} = h_{ab}dx^{a}dx^{b} + R^{2}(t)(d\theta^{2}+\sin^{2}\theta d\varphi^{2}),
\end{equation}
where $R(t,r)$ is the physical radius of the FLRW universe, $R(t,r):=a(t)r$, being $r$ the co-moving coordinate and $a(t)$ the scale factor. Besides, we have defined the following metric, $h_{ab} := \mbox{diag}[-1,a^{2}(t)/(1-kr^{2})]$, in this case $a,b=0,1=t,r$. An interesting property of this spacetime is the existence of a marginally trapped null surface, known as apparent horizon, and can be obtained from the following condition
\begin{equation}
    h^{ab}\partial_{a}R \partial_{b}R =0,
\end{equation}
which leads to the following expression
\begin{equation}
    R_{A}=\frac{1}{\sqrt{H^{2}+\frac{k}{a^{2}}}},
\end{equation}
being $H$ the Hubble parameter. As can be seen, the above expression determines the apparent horizon as a dynamic structure. It is also argued that the apparent horizon of a dynamic system can be interpreted as a causal horizon. We can associate surface gravity and gravitational entropy with it. In this case, the surface gravity is given by
\begin{equation}
    \kappa = \frac{1}{2\sqrt{-h}}\partial_{a}(\sqrt{-h}h^{ab}\partial_{b}R),
\end{equation}
the explicit expression for the above equation takes the form
\begin{equation}
    \kappa = -\frac{R}{2}\left[\dot{H}+2H^{2}+\frac{k}{a^{2}} \right],
\end{equation}
where the dot denotes derivatives w.r.t. cosmic time $t$. Besides, if the apparent horizon is considered as a thermodynamic system, we can define the temperature as 
\begin{equation}
    T_{A} = \frac{\kappa}{2\pi}.
\end{equation}
which is positive for an expanding cosmology. As mentioned before, in references \cite{helou, posit} the nature of the horizon and its corresponding temperature were studied for several situations. For spherically symmetric apparent horizons, the first law of thermodynamics can be written in the form
\begin{equation}
    dE = A\Psi + WdV,
\end{equation}
this expression is usually termed as unified first law, $A$ denotes the area of a sphere of radius $R$ and $V$ the volume, $\Psi$ is the energy flux defined as follows 
\begin{equation}
    \Psi = \Psi_{a}dx^{a} = (T_{a}{}^{b}\partial_{b}R+W \partial_{a}R)dx^{a}.
\end{equation}
In this case, $T_{ab}$ is the energy-momentum tensor of the perfect fluid describing the content of the universe, $T_{ab} = (\rho+p)u_{a}u_{b}+pg_{ab}$ and $W$ is the work density of the matter fields defined as
\begin{equation}
    W=-\frac{1}{2}T = \frac{1}{2}(\rho-p),\label{eq:work}
\end{equation}
being $T$ the two-dimensional normal trace of the energy-momentum tensor, $T:=h^{ab}T_{ab}$. Evaluating the above expressions, one gets the energy supply vector 
\begin{equation}
    \Psi_{a} = \left(-\frac{1}{2}(\rho+p)HR, \frac{1}{2}(\rho+p)a \right),
\end{equation}
which leads to the following equation
\begin{equation}
    dE=2\pi R^{2}(\rho+p)(-HRdt+adr)+\frac{1}{2}(\rho-p)dV. \label{eq:diffe}
\end{equation}
Once we evaluate at the apparent horizon the above results, we can interpret $\Psi_{a}$ as the total energy flow at the horizon, and $W$ can be viewed as the work done by the change of the apparent horizon. This quantity can be related to the thermodynamic pressure, $P$ since it can be considered as the conjugate variable of the thermodynamic volume from the first law of thermodynamics; we have $W:=P$.

\subsection{$P-v$ phase transitions}\label{pv}
If we specialize in the case $k=0$ and evaluate at $R=R_{A}$, we can write for the apparent horizon the following quantities
\begin{equation}
    R_{A} =\frac{1}{H}, \label{eq:ahknull}
\end{equation}
therefore
\begin{equation}
    \kappa = -\frac{1}{R_{A}}\left( 1-\frac{\dot{R}_{A}}{2HR_{A}}\right),
\end{equation}
and in consequence, we can write the positive temperature for the apparent horizon as
\begin{equation}
    T_{A} = \frac{1}{2\pi R_{A}}\left( 1-\frac{\dot{R}_{A}}{2HR_{A}}\right) = \frac{1}{4\pi R_{A}}(1-q).\label{eq:temp}
\end{equation}
It is worth mentioning that the temperature depends on the $R_{A}$ and its change along cosmic evolution, the case $\dot{R}_{A}/(2HR_{A})\ll 1$ in the latter expression resembles the temperature of a spherically symmetric black hole with horizon radius $R_{A}$. In general, the temperature on the horizon is defined by the characteristics of the cosmic expansion generated by the energy content of the universe through the deceleration parameter $q$, which is defined as $1+q := -\dot{H}/H^{2}$. This expression is useful to notice that the positive of the temperature is maintained for quintessence and phantom dark energy.\\ 

The Friedmann constraint describes the background dynamics for two non-interacting fluids in an FLRW spacetime as follows
\begin{equation}
    3H^{2} = 8\pi (\rho_{\mathrm{m}}+\rho_{\mathrm{de}})=8\pi \rho_{de} (1+r),\label{eq:fried1}
\end{equation}
where $\rho_{\mathrm{m}}$ characterizes the energy density of the dark matter and $\rho_{\mathrm{de}}$ is associated to the dark energy sector. We have defined the cosmological coincidence parameter, $r$, given as the quotient $r:=\rho_{\mathrm{m}}/\rho_{\mathrm{de}}$. The acceleration equation is written in the usual form
\begin{equation}
    \dot{H} = -4\pi (\rho_{\mathrm{m}}+\rho_{\mathrm{de}}+p_{\mathrm{m}}+p_{\mathrm{de}}). \label{eq:fried2}
\end{equation}
Using the Friedmann equations (\ref{eq:fried1}), (\ref{eq:fried2}) and the apparent horizon definition given in (\ref{eq:ahknull}), then one writes 
\begin{equation}
    \rho_{\mathrm{de}} = \frac{3}{8\pi (1+r) R^{2}_{A}},\label{eq:derestrict}
\end{equation}
and
\begin{equation}
    p_{\mathrm{m}}+p_{\mathrm{de}} = \frac{\dot{R}_{A}}{4\pi R^{2}_{A}}-\frac{3}{8\pi R^{2}_{A}}.
\end{equation}
Therefore, the work density (\ref{eq:work}) takes the form 
\begin{equation}
    P=\frac{T_{A}}{2R_{A}}+\left(\frac{3}{1+r}-1\right)\frac{1}{16\pi R^{2}_{A}},\label{eq:eos1}
\end{equation}
where a barotropic equation of state of the form, $p=\omega \rho$, was considered for each specie and $\omega_{\mathrm{m}} = 0$ to describe a pressureless fluid for the matter sector. Besides, the Eq. (\ref{eq:temp}) was used to substitute $\dot{R}_{A}$ by $T_{A}$. Consider the usual definition for the specific volume, $v=2R_{A}$. The latter expression is the thermodynamic equation of state for the FLRW universe in which we consider two non-interacting fluids to describe the universe's matter content. The conditions to have a $P-v$ phase transition in the system are 
\begin{equation}
    \left(\frac{\partial P}{\partial v} \right)_{T_{A}} = \left(\frac{\partial^{2}P}{\partial v^{2}} \right)_{T_{A}} = 0, \label{eq:crit2}
\end{equation}
and when applied to (\ref{eq:eos1}), no critical points are found; therefore, the model does not admit $P-v$ phase transitions. 
Notice that the lack of critical points in the coincidence parameter is written in terms of the redshift or scale factor. Then, the coincidence parameter must be related to the specific volume $v$ (or $R_{A}$); this is crucial to allow $P-v$ phase transitions in Einstein gravity. According to Eq. (\ref{eq:derestrict}), the demanding condition discussed above on $r$ to allow phase transitions in our scheme severely restricts the form of the energy density associated with the dark energy sector. However, if we adopt the usual holographic dark energy formula, we can write directly the desired expression, $r=r(R_{A})$. This represents an interesting scenario since, as discussed in \cite{colgain}, the holographic principle is a propitious arena to confront theory, model assumptions, and recent observations. The holographic principle is the most important cornerstone of quantum gravity and allows us to describe the physical quantities inside the universe, including the energy density of the dark energy sector, by some quantities on the universe's boundary. Based on the dimensional analysis, only two quantities can be used to construct the dark energy density: the reduced Planck mass, $M_{P}$, and the characteristic length $L$ \cite{hp, Nojiri:2005pu, Nojiri:2019yzg}
\begin{equation}
    \rho_{\mathrm{de}}= C_{1}M^{4}_{P}+C_{2}M^{2}_{P}L^{-2}+C_{3}L^{-4}+\mathcal{O}(L^{-6})+...
    \label{eq:expansion}
\end{equation}
where the parameters $C_{1}, C_{2}, C_{3}$ are constant. The $C_{1}$ term is not compatible with the holographic principle, it is well established that such a term leads to the cosmological constant problem, then the expansion given above should start from the second term\footnote{It is important to stress out that the form of (\ref{eq:expansion}) comes from the application of the renormalization group method in cosmology, i.e., the full vacuum energy is a physical observable when the leading quantum contributions to it are characterized, such contributions are given by an even-power law of the Hubble expansion rate, $\rho_{\Lambda}(H) \propto \rho^{0}_{\Lambda}+M^{2}_{P}H^{2}+\mathcal{O}(H^{4})+\mathcal{O}(H^{6})+...$, where $\rho_{\Lambda}$ is the renormalized part of the complete QFT structure of the vacuum energy. For a complete discussion, see Ref. \cite{Nojiri:2017opc, qft} and references therein.}. According to our scheme, the Hubble scale gives the most convenient election for the characteristic length, which is defined as $L:=1/H$. For generalization purposes, we used the spirit proposed by \cite{Nojiri:2017opc}, where the authors introduced a generalized holographic dark energy model of the form
\begin{equation}
L=L(L_{p},\dot{L}_{p}, \ddot{L}_{p},...,L_{f},\dot{L}_{f}, \ddot{L}_{f},...,H, \dot{H},\ddot{H},...,t_{s}).\label{eq:gene}    
\end{equation}
The infrared Nojiri-Odintsov cutoff corresponds to a combination of the FRW universe parameters: the Hubble parameter, particle ($L_{p}$) and future horizons ($L_{f}$), and their derivatives, cosmological constant and the universe lifetime, $t_{s}$, (if finite). In our new approach to finding and describing the $P-v$ phase transition, we use the simplest choice of the cutoff (\ref{eq:gene}), which corresponds to $L=L(H)$ (and its generalization $L=L(H,\dot{H}$)) because it is possible to map such holographic dark energy to modified gravity or gravity with general fluid. Therefore, we consider the following form for the energy density of the dark energy sector
\begin{equation}
    \rho_{\mathrm{de}} =3\beta H^{2n}, \label{eq:rho}
\end{equation}
where $\beta$ and $n$ are real parameters with $n \geq 1$ being an integer. The existence of possible analytic solutions for the critical points equations (\ref{eq:crit2}) remain in the form that we choose the holographic dark energy, Eq. (\ref{eq:rho}) in our case. The dark energy model (\ref{eq:rho}) can be related to the geometry of the apparent horizon through the specific volume $v$ (or $R_A$) in a simple way due to Eq. (\ref{eq:ahknull}), in consequence the coincidence parameter acquires the desired functional form, $r=r(v)$; this is the critical ingredient to characterize in an analytic path the $P-v$ phase transitions in Einstein's gravity, as commented above for the critical conditions (\ref{eq:crit2}), notice that the form given above for the energy density covers some of the terms given in the expansion (\ref{eq:expansion}), we will focus on some cases with physical meaning, as we will discuss below. The classical stability of the dark energy models emerging from Eq. \ref{eq:rho} is discussed in the appendix \ref{sec:app}. Inserting Eq. (\ref{eq:rho}) in (\ref{eq:derestrict}) we can write the following expression
\begin{equation}
    1+r=\frac{1}{8\pi \beta}R_{A}^{2(n-1)},
\end{equation}
where we have also considered (\ref{eq:ahknull}). In this case the Eq. (\ref{eq:eos1}) takes the form
\begin{equation}
    P=\frac{T_{A}}{2R_{A}}+\left(\frac{24\pi \beta}{R^{2(n-1)}_{A}}-1\right)\frac{1}{16\pi R^{2}_{A}}.\label{eq:eos2}
\end{equation}
Notice that for $n=1$, the second term in the last expression is constant. Therefore, no critical points can be found in this case since it resembles the usual equation of state (\ref{eq:usual}). In terms of the specific volume defined above, for $n>1$ we obtain from the conditions (\ref{eq:crit2}), the following critical point
\begin{equation}
    v_{c}=\left[2^{2n} 6\pi \beta n (2n-1) \right]^{1/2(n-1)},\label{eq:point1}
\end{equation}
and
\begin{equation}
    T_{c} = \frac{1}{\pi v_{c}}\left(\frac{n-1}{2n-1} \right), \label{eq:point2}
\end{equation}
which is positive. Evaluating (\ref{eq:eos2}) at the critical values of the specific volume and temperature one gets
\begin{equation}
    P_{c}=\frac{1}{\pi v_{c}^{2}}\left(\frac{n-1}{2n-1} \right)+\left(\frac{2^{2n}6 \pi \beta}{v_{c}^{2(n-1)}}-1\right)\frac{1}{4\pi v^{2}_{c}}. \label{eq:point3}
\end{equation}
Therefore for $n>1$ the model admits $P-v$ phase transitions. It is worth mentioning that in the context of holographic dark energy $n=1$ in Eq. (\ref{eq:rho}) represents the model proposed by Cohen in Ref. \cite{cohen}. However, as pointed out by Hsu and Li \cite{Hsu:2004ri, li}, this model is not viable as a dark energy candidate since the dark energy scales with the universe scale factor, $a$, in the same way as dark matter, i.e., $a^{-3}$. The dark energy is a pressureless fluid with $\omega=0$, which is incompatible with current observations. The unfeasibility for the phase transition in this model is exhibited in our analysis by obtaining a constant behavior for the cosmological coincidence parameter.\\ 

In the following, we will focus on specific values for $n$ that are physically justified within the cosmological context. The value $n=2$ corresponds to a dark energy model recently proposed in Refs. \cite{transient1, transient2} with the following important properties: the dark energy parameter state can be written in terms of the coincidence parameter defined as, $r:=\rho_{\mathrm{m}}/\rho_{\mathrm{de}}$, yielding 
\begin{equation}
    \omega_{\mathrm{de}} = -1 -2(1+\omega_{\mathrm{m}})\frac{r(z)}{1-r(z)},
\end{equation}
where the coincidence parameter in terms of the redshift, $z$, has the explicit form
\begin{equation}
    r(z)=\frac{\lambda(1+z)^{3}}{\left(1+\sqrt{1-\lambda(1+z)^{3}} \right)^{2}},
    \label{eq:coincidence}
\end{equation}
being $\lambda$ a constant parameter defined as, $\lambda:=4[1-\Omega_{\mathrm{m,0}}]\Omega_{\mathrm{m,0}}$, where $\Omega_{\mathrm{m,0}}$ is the fractional energy density associated to the matter sector at present time; then the model describes an early phantom scenario since at the beginning of the dark energy dominance epoch given at $r=1$, the parameter state $\omega_{\mathrm{de}}$ diverges. In the case of the future evolution of these dark energy models differs considerably despite the similarities between ghost dark energy and generalized holographic dark energy; the full classification for future type V singularities can be seen in Ref. \cite{deHaro:2023lbq, Trivedi:2023zlf, Trivedi:2022ngt}. However, its energy density remains bounded, thus the singular behavior is obtained only in the parameter state (pressure) of the fluid. As can be seen in Ref. \cite{transient1, transient2}  previously, our scheme admits a phantom scenario whose singularity lies only on the parameter state of dark energy and appears in the deceleration-acceleration transition stage; from there, the model tends to a future de Sitter evolution, i.e., our approach allows a transient early (not future) phantom regime. A signal of stability for this dark energy model is obtained from the positivity of the squared adiabatic sound speed.
In fact, at the far future ($z=-1$) from Eq. (\ref{eq:coincidence}) we can observe that $\omega_{\mathrm{de}}(z\rightarrow -1)\rightarrow -1$, therefore the model exhibits a future transition  to de Sitter expansion, further details of the model are discussed in \cite{transient1, transient2}. Beyond Einstein's gravity, it is worth mentioning that a $H^{4}$ term also appears in the framework of cubic gravity\footnote{In this scenario, the Einstein-Hilbert action is corrected by the term $P$ as follows
\begin{equation}
    S=\int d^{4}x\sqrt{-g}\left[\frac{1}{2\kappa}(R-2\Lambda)+\alpha P \right],\label{eq:cubic}
\end{equation}
where $P$ is a general non-topological cubic term that involves curvature invariants.} as a correction term for cosmic evolution \cite{cubic}. The case $n=3$ will also be considered, a generalization of the type, $P \rightarrow f(P)$ in action (\ref{eq:cubic}), being $f$ an arbitrary function of the cubic term, admits correction terms of the form $H^{6}$ in the acceleration equation \cite{sixth}, in this kind of model the parameter state of the dark energy sector also experiences a future transition from quintessence (phantom) to de Sitter evolution, the behavior of dark energy during the cosmological evolution, in this case, depends on the election of the function $f$. Finally, the cosmological evolution of quartic gravity was explored in Ref. \cite{oliva} and corrected by a term of the form $H^{8}$, which corresponds to the case $n=4$; for this scenario, was found that a matter dominated epoch is smoothly connected to a late time acceleration described by a de Sitter phase. It is worth mentioning that gravitational theories discussed above represent curvature corrections to General Relativity whose vacuum spectrum consists of a graviton and is ghost-free \cite{cubic, sixth, oliva}, a relevant characteristic of these modified scenarios is that contrary to what turns out to be a generic feature of higher curvature contributions, the Friedmann equations preserve their second order in the derivatives, this feature avoids the presence of ill behaviors as ghost or massive modes. Within the cosmological scenario, only powers of $H$ are expected to contribute to the critical density in order to maintain the aforementioned properties of the theory; for details, see \cite{cubic, sixth, oliva}. Thus, we consider two principal arguments to construct our energy density for the dark energy sector: (i) the simplest form of the generalized holographic cutoff given in Eq. (\ref{eq:gene}) and (ii) in order to avoid the propagation of extra degrees of freedom which is guaranteed by a set of second order Friedmann equations, makes dark energy as the one given in Eq. (\ref{eq:rho}), a suitable election to contribute to the energy density of the universe.

Additionally, powers of the Hubble parameter in the Friedmann equations are usually found in various theories of gravity, including Lovelock and scalar-tensor theories, such as Refs. given in \cite{h6, various}. 

Using the equations (\ref{eq:point1}), (\ref{eq:point2}) and (\ref{eq:point3}) we can write
\begin{equation}
    \frac{P_{c}v_{c}}{T_{c}}=1+\frac{1}{4}\left[\frac{1-n(2n-1)}{n(n-1)} \right],
\end{equation}
which is a constant value and independent of the parameter $\beta$. For $n=1$ the latter expression diverges, therefore $P-v$ phase transitions are expected for $n>1$. On the other hand, if we specialize the above expression for the values $n=2,3,4$ we obtain $3/8, (10/9)(3/8), (7/6)(3/8)$, respectively. Notice that the constant obtained in each case is proportional to the universal ratio obtained from the van der Waals theory given by $3/8$, which establishes an upper bound for real gases. For the fluids discussed here, this value represents a lower bound for the ratio $(P_{c}v_{c})/T_{c}$. This indicates that for $3/8$, the existence of phase transitions in these cosmological fluids is consistent with the mean-field theory description, and in the cases where deviations from $3/8$ are obtained, the phase transitions can be due to interactions between the constituents of the fluid beyond the van der Waals gas description \cite{vanderwaals}.   

\begin{figure}[htbp!]
\centering
\includegraphics[scale=0.68]{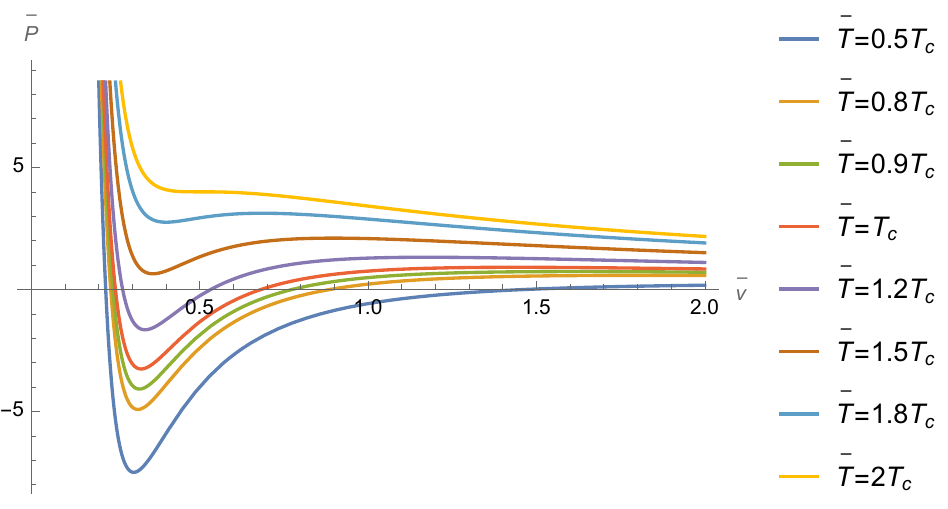}
\caption{$\bar{P}-\bar{v}$ isotherms for the FLRW universe with $H^{4}$ dark energy.}
\label{fig:pv}
\end{figure}

To have dimensionless quantities, we define the following reduced variables: $\bar{P}:=P/P_{c}$, $\bar{T}:=T/T_{c}$ and $\bar{v}:=v/v_{c}$, thus Eq. (\ref{eq:eos2}) can be written as follows
\begin{equation}
    \bar{P}=\frac{\bar{T}}{\bar{v}}\frac{T_{c}}{v_{c}P_{c}}+\left(\frac{2^{2n}6 \pi \beta}{v_{c}^{2(n-1)}}\frac{1}{\bar{v}^{2(n-1)}}-1\right)\frac{1}{4\pi \bar{v}^{2}}\frac{1}{P_{c}v^{2}_{c}}.\label{eq:crit3}
\end{equation}
In Fig. (\ref{fig:pv}), we show the behavior of expression (\ref{eq:crit3}) for the case $n=2$. As seen in the plots, the isotherms can exhibit a coexistence phase. The shape of the isotherms is not affected by the parameter $\beta$ value. Notice that the $P-v$ diagram is quite similar to the one obtained for the van der Waals gas.\\ 

\begin{figure}[htbp!]
\centering
\includegraphics[scale=0.68]{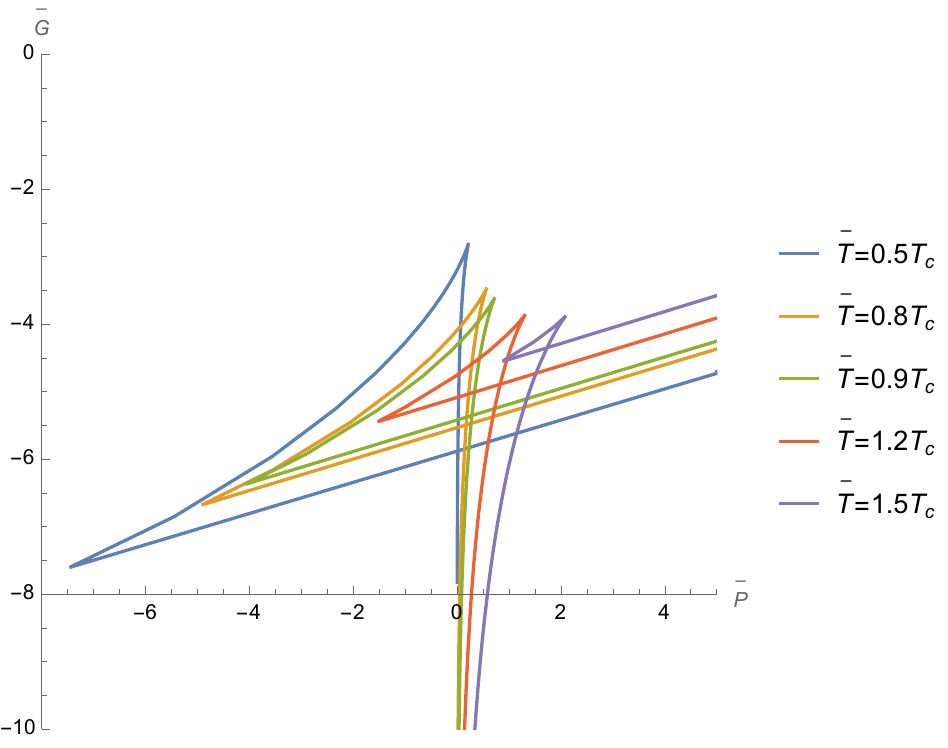}
\caption{Reduced Gibbs free energy against reduced pressure for $n=2$.}
\label{fig:gibbs1}
\end{figure}

\begin{figure}[htbp!]
\centering
\includegraphics[scale=0.48]{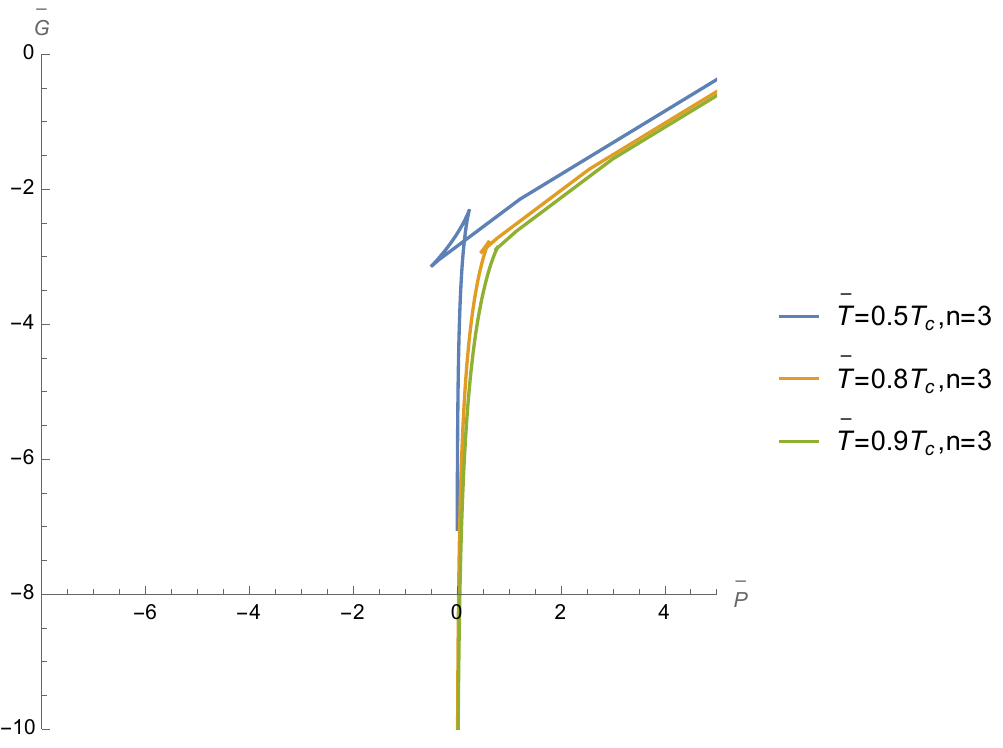}
\includegraphics[scale=0.48]{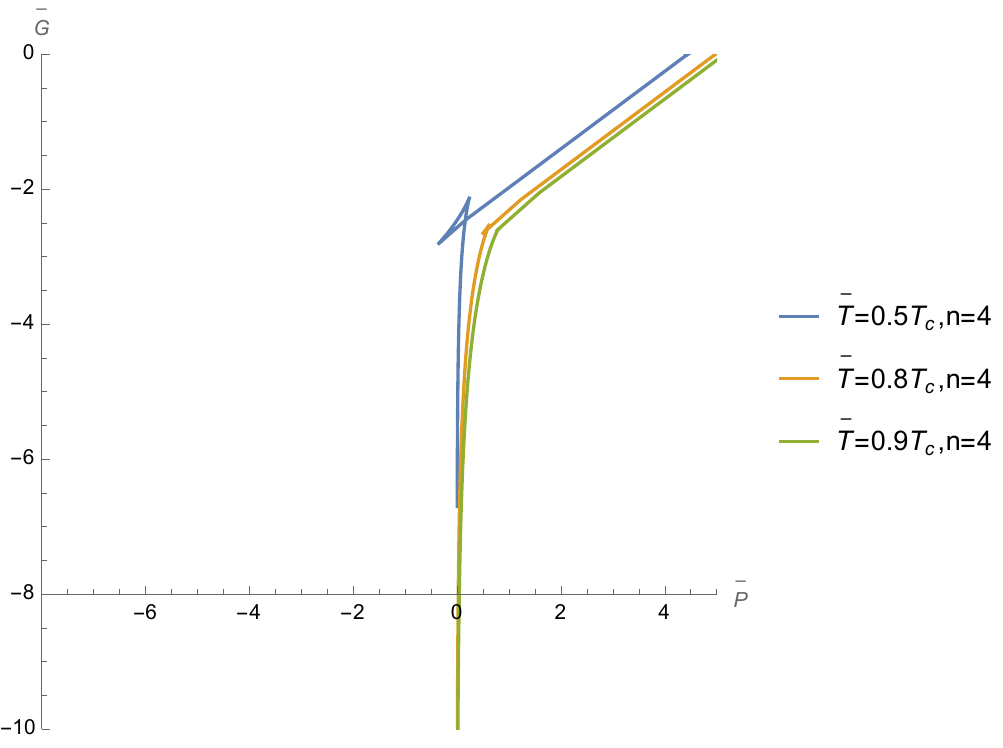}
\caption{Reduced Gibbs free energy against reduced pressure for $n=3$ and $n=4$.}
\label{fig:gibbs2}
\end{figure}

Using the plot for the reduced Gibbs free energy against the reduced pressure, $\bar{P}$, we can characterize the phase transition observed in the $P-v$ diagram given above. Fig. (\ref{fig:gibbs1}) depicts the reduced Gibbs free energy behavior. We obtain $\bar{G}$ from the standard thermodynamic relation
\begin{equation}
    \left(\frac{\partial \bar{G}}{\partial \bar{v}} \right)_{\bar{T}} = \bar{v}\left(\frac{\partial \bar{P}}{\partial \bar{v}} \right)_{\bar{T}}.
\end{equation}
According to figure (\ref{fig:gibbs1}) we observe the typical shape that characterizes a first-order phase transition; all unstable states of the system are contained in the loop of the plot whose size is modified by the temperature value. We also observe first-order phase transitions for cases $n=3$ and $n=4$. However, contrary to the case $n=2$, the isotherms obtained above the critical temperature value only exhibit monotonically decreasing behavior. In Fig. (\ref{fig:gibbs2}), we show the reduced Gibbs free energy behavior for $n=3$ and $n=4$. 
Notice that the size of the loop characterizing the first-order phase transition decreases for two different situations: (i) different dark energy models as the value of $n$ increases and (ii) for a fixed $n$ but as we increase the value of the temperature. The first case indicates that for higher powers in $H$, the dark energy models proposed here are more stable thermodynamically speaking, and in consequence, the $P-v$ diagram in such cases will be more similar to the one associated with the ideal gas. Hence, no phase transition will be observed. Aside from the physical justification from diverse cosmological models, the thermodynamic description imposes important bounds on the values of the exponent $n$ that can be considered in the dark energy density (\ref{eq:rho}) to obtain phase transitions in the model. As can be observed in Fig. (\ref{fig:pv}), the coexistence phase per isotherm in the $P-v$ diagram is obtained for small values of the specific volume since we have $v \propto H^{-1}$, the phase transition takes place at the future. This means that phase transitions in the thermodynamic context reveal the transient behavior of dark energy and its passage from phantom (quintessence) to a future de Sitter expansion, as discussed before for the cosmological models considered here. It is of the utmost importance that thermodynamic phase transitions occur in the future evolution. If a radical phase transition had occurred in the recent past, it would have left observable imprints \cite{diaz}.\\

Based on the above-mentioned thermodynamic construction and the generalized holographic dark energy models presented in \cite{Nojiri:2017opc}, we extend our formalism developed to holographic dark energy models, which contain a linear term in the first derivative of the Hubble parameter. We use the following modification for the energy density, (\ref{eq:rho})
\begin{equation}
        \rho_{\mathrm{de}} =3(\beta H^{2n}+\gamma \dot{H}), \label{eq:rho2}
\end{equation}
with $\beta$ and $\gamma$ being positive arbitrary parameters; for $n=1$, the Granda-Oliveros holographic model is recovered \cite{grandaoliveros}, which in turn results to be an extension of the Ricci holographic model \cite{ricci}, see also \cite{ricci2} for an alternative extension of this model. Notice that a generalization of this type is permitted since the first derivative of $H$ is related to the temperature of the horizon through Eq. (\ref{eq:temp}). The inclusion of the first derivative of the Hubble parameter into the dark energy density can also be motivated with scenarios beyond the holographic approach. In the context of QCD, minimal extensions of the {\it ghost dark energy} consider the first derivative of $H$. In this scenario, the magnitude of the vacuum energy is corrected by the contribution of some ghost fields given as $\rho \propto H \Lambda^{3}_{QCD}$, being $\Lambda_{QCD}$ the QCD mass scale. See, for instance, the references given in \cite{qcd} for a complete guide on the topic and the cases studied when the first derivative of $H$ is considered as a correction term to the ghost dark energy.\\ 

\begin{figure}[htbp!]
\centering
\includegraphics[scale=0.68]{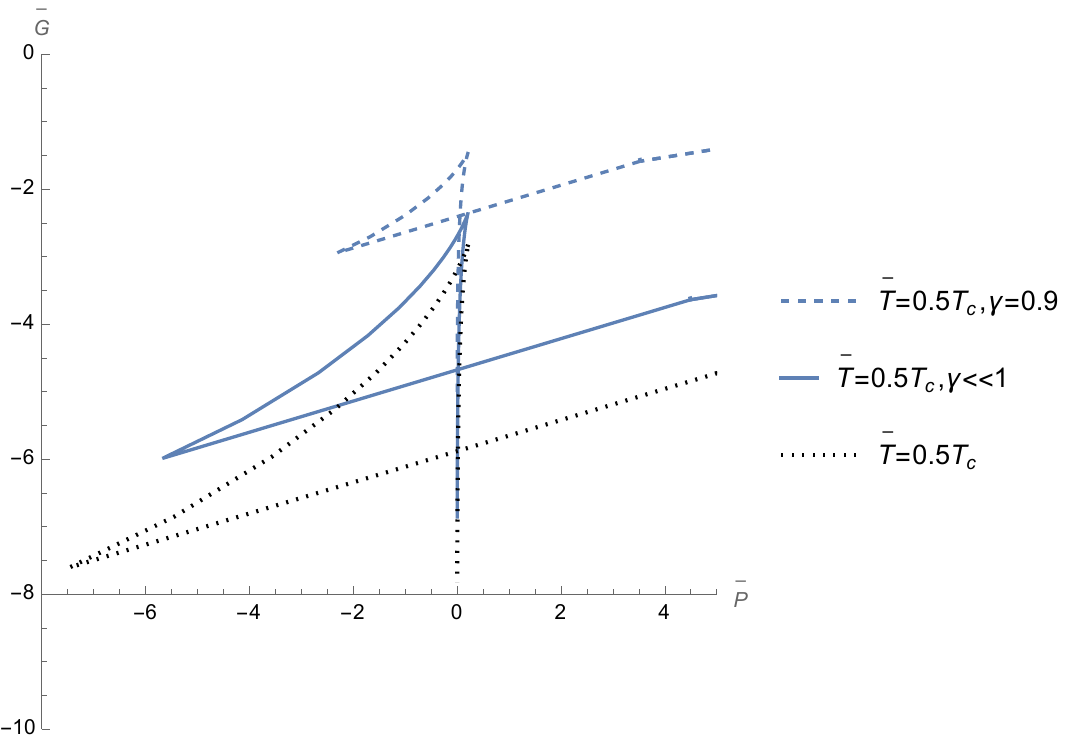}
\caption{Reduced Gibbs free energy against reduced pressure for energy density (\ref{eq:rho}) (dotted line) and (\ref{eq:rho2}) (solid and dashed lines).}
\label{fig:gibbs3}
\end{figure}

If we repeat the procedure exposed before and use the expression of the temperature (\ref{eq:temp}) to relate $T_{A}$ with $\dot{R}_{A}$, we find the following expression for the thermodynamic equation of state  
\begin{equation}
    P=\frac{T_{A}}{2R_{A}}(1+12\pi \gamma)+\left(\frac{24\pi \beta}{R^{2(n-1)}_{A}}-48\pi \gamma - 1\right)\frac{1}{16\pi R^{2}_{A}}.\label{eq:eos3}
\end{equation}
Notice that for $\gamma=0$ we recover the Eq. (\ref{eq:eos2}). We can observe that for $n=1$, we obtain a constant value in the denominator of $24\pi \beta$; this behavior leads to no phase transition. Once again, the $P-v$ phase transitions are expected to be obtained for $n>1$. If we consider this latter case for $n$ one gets the critical point with a positive specific volume given by
\begin{equation}
    v_{c}=\left[\frac{2^{2n}6\pi \beta n}{1+48\pi \gamma}(2n-1) \right]^{1/2(n-1)},
\end{equation}
the critical temperature can be written as
\begin{equation}
    T_{c} = \frac{1}{\pi v_{c}(1+12\pi \gamma)}\left(\frac{n-1}{2n-1} + 24 \pi \gamma \right), 
\end{equation}
and finally, the critical pressure takes the form
\begin{equation}
    P_{c}=\frac{1}{\pi v_{c}^{2}}\left(\frac{n-1}{2n-1} +24\pi \gamma \right)+\left(\frac{2^{2n}6 \pi \beta}{v_{c}^{2(n-1)}}-48 \pi \gamma - 1\right)\frac{1}{4\pi v^{2}_{c}}.
\end{equation}
The inclusion of $\dot{H}$ into the dark energy density is identified by a constant term mediated by the $\gamma$ parameter in the critical solution. In terms of the reduced variables, we can write Eq. (\ref{eq:eos3}) as follows
\begin{equation}
    \bar{P}=\frac{\bar{T}}{\bar{v}}\frac{T_{c}}{v_{c}P_{c}}(1+12\pi \gamma)+\left(\frac{2^{2n}6 \pi \beta}{v_{c}^{2(n-1)}}\frac{1}{\bar{v}^{2(n-1)}} - 48\pi \gamma - 1\right)\frac{1}{4\pi \bar{v}^{2}}\frac{1}{P_{c}v^{2}_{c}}.\label{eq:eos3a}
\end{equation}
For comparative purposes between (\ref{eq:rho2}) and (\ref{eq:rho}), we focus only on the behavior of the reduced Gibbs free energy in the case $n=2$. In Fig. (\ref{fig:gibbs3}), we show such behavior; as can be seen, the value of the concomitant parameter to $\dot{H}$, namely $\gamma$, modifies the size of the loop that characterizes the first-order phase transition; as its value increases, we observe that the loop's size decreases. Then, in this scenario, the thermodynamic instability can be attenuated, which means that the isotherms obtained in the $P-v$ diagram are smoothed. Similar to the energy density (\ref{eq:rho}) results, the outcomes emerging from (\ref{eq:rho2}) are independent of the parameter $\beta$.     

\section{Conclusions}\label{conclusion_f}

The aim of this work was to demonstrate the existence of $P-V$ phase transitions within the thermodynamic description of Einstein gravity for the case of a flat FLRW spacetime; as far as we know, this is the first work in which this result is established. We focused our attention on cosmic evolution in the late times. Therefore, the universe's content is described by two dominant fluids: dark matter and dark energy. In our scheme, the fluids of the dark sector are not allowed to interact but are related throughout the usual cosmological coincidence parameter, which we denote by $r$. Since the parameter $r$ is, in general, expected to be a dynamical function, we must be able to write it in terms of the geometry of the apparent horizon, i.e., it must be a function of the radius, $R_{A}$, which is the dynamical quantity for an expanding cosmology. Considering two fluids, this is a compelling requirement for the parameter $r$ to guarantee the existence of $P-V$ phase transitions in our approach. Furthermore, the prerequisite mentioned above for $r$ is backed by the conditions that establish the existence of critical points in the equation of state of the universe, namely $P=P(V,T)$; such conditions are defined through partial derivatives w.r.t. $R_{A}$. 

According to our construction, the form of the dark energy density is restricted to be proportional to a power-law function of the Hubble parameter, $H$. We restrict ourselves to values of the exponent $n$ that are physically justified. For instance, $n=1$ represents the well-known holographic dark energy widely explored in the literature with characteristic length given by the Hubble scale. The case $n=2$ is a dark energy model that leads to an early phantom scenario in Einstein's gravity, and the cases $n=3,4$ resemble the correction terms appearing in the acceleration equation of the cosmic evolution emerging from modified gravity theories in 4D and those in which cubic and quartic invariants of the curvature tensor are taken into account. However, other holographic models can be easily adapted to our analysis, see for instance \cite{hdemodels} where the Barrow, Tsallis, Renyi, and Sharma-Mittal (to mention some) are dark energy models which fulfill the condition, $\rho_{\mathrm{de}} \propto H$. Therefore, this simple assumption on the form of the dark energy density can encompass several cases of interest. We also extended our analysis to the case in which a linear term modifies the dark energy density in the first derivative of $H$. This contribution is properly studied within the holographic dark energy scenario (Ricci and Granda-Oliveros models). In QCD, this type of modification appeared in the context of the so-called {\it ghost dark energy}.

For $n=1$, no critical points are found; consequently, no $P-V$ phase transitions are observed. We attribute this result to the cosmological coincidence parameter exhibiting a constant behavior. In this case, the $P(V,T)$ equation of state does not differ from the usual expression discussed in Einstein's gravity. However, for $n>1$, the isotherms displayed in the $P-V$ diagram show the coexistence phase for temperature values below the critical temperature, and the isotherms over the critical temperature indicate an ideal gas behavior as the temperature increases. Therefore, a $P-V$ phase transition is expected for the cosmological fluid, and according to the Gibbs free energy behavior, we have a first-order phase transition. On the other hand, as the value of the exponent $n$ increases we observe that the thermodynamic instability is attenuated, in other words, the size of the loop characterizing the first-order transition decreases, this means that in the $P-V$ diagram the isotherms that show coexistence phase are smoothed. Therefore, higher powers in $H$ could lead to scenarios with no phase transitions, and this characteristic provides a cutoff in the value of the exponent $n$. An interesting result is that the thermodynamic description does not depend on the value of the arbitrary parameter introduced in the dark energy density. The variations obtained between different cases are due to the value of the exponent $n$ only.

In the extended case, we observe that the first derivative of $H$ could lead to scenarios with ideal gas behavior as the value of its coupling parameter, denoted by $\gamma$, increases. 

Finally, we emphasize that there are still a lot of subtle issues worth investigating when tests against observations are implemented for the holographic dark energy models treated in this work, for instance, the restriction of the exponent $n$ appearing in (\ref{eq:rho}) with astrophysical data. However, we feel that any further exploration along this line would justify a separate analysis.

\begin{acknowledgments}
M. Cruz work has been supported by S.N.I.I. (CONAHCyT-M\'exico). J. Saavedra acknowledges the financial support of Fondecyt Grant 1220065. M. Cruz also acknowledges the warm hospitality of the Physics Institute of PUCV and valuable discussions with F. Tello-Ortiz during the preparation of this manuscript. Anonymous reviewers are acknowledged for helpful suggestions.
\end{acknowledgments}

\appendix
\section{Classical stability}
\label{sec:app}
Following the line of reasoning of Ref. \cite{stability}, we study the stability of the background evolution of the dark energy models arising from Eq. (\ref{eq:rho}) for $n=1,2,3,4$ by analyzing the sign of the squared speed of sound given by 
\begin{equation}
    v^{2} = \frac{dp_{\mathrm{de}}}{d\rho_{\mathrm{de}}},
\end{equation}
where $v^{2} < 0$ represents classical instability of the model at perturbation level. For a barotropic equation of state the above equation is simply
\begin{equation}
    v^{2} = \omega_{\mathrm{de}}+\rho_{\mathrm{de}}\frac{d\omega_{\mathrm{de}}}{d\rho_{\mathrm{de}}}.\label{eq:vel}
\end{equation}
Inserting the dark energy (\ref{eq:rho}) in the continuity equation $\dot{\rho}_{\mathrm{de}} + 3H\rho_{\mathrm{de}} (1+\omega_{\mathrm{de}})=0$, one gets
\begin{equation}
    \omega_{\mathrm{de}} = -1+\frac{2n}{3}(1+q) = -1 + 2n\left(\frac{\rho_{\mathrm{m}}}{2\rho_{\mathrm{m}}-(2n-1)\rho_{\mathrm{de}}}\right) ,
\end{equation}
where we have used the Friedmann constraint (\ref{eq:fried1}) to write the second equality. Therefore we can write Eq. (\ref{eq:vel}) as follows
\begin{equation}
    v^{2} = \frac{4n}{3-2n}\left(\frac{r^{2}(z)}{2r(z)-2n+1} \right),
\end{equation}
being $r(z)$ the coincidence parameter defined previously. Notice that for the epoch of dark energy dominance the coincidence parameter is restricted to the interval $(0,1]$ where $r\approx 0$ at the future ($z < 0$) and $r\approx 1$ at early times ($z < 1$) as well as present time ($z=0$); thus for $n=1$ we have $v^{2} < 0$ and for $n=2,3,4$ we obtain $v^{2} > 0$ during early and present times cosmic evolution, respectively. According to our results the case $n=1$ it lacks of physical interest since no $P-v$ phase transition is obtained and in turn results to be classically unstable.
Regarding the model given by Eq. (\ref{eq:rho2}) we follow the discussion given in Ref. \cite{unstable}. In general, the consideration of the term $\dot{H}$ induces instabilities in the holographic dark energy model. However, the introduction of more parameters into the holographic energy density can lead to stable scenarios. We consider that in our case different values for the exponent $n$ could have the same effect as the extra parameter (Barrow exponent) considered and discussed in \cite{unstable}. 



\begin{thebibliography}{99} 
\bibitem{bekenstein}
J.~D.~Bekenstein, Lett.\ Nuovo\ Cimento {\bf 4}, 737 (1972); Phys.\ Rev.\ D {\bf 7}, 2333 (1973); Phys.\ Rev.\ D {\bf 9}, 3292 (1974). 

\bibitem{haw}
S.~W.~Hawking, Phys.\ Rev.\ Lett. {\bf 26}, 1344 (1971).

\bibitem{haw2}
G.~W.~Gibbons and S.~W.~Hawking, Phys.\ Rev.\ D {\bf 15}, 2738 (1977).

\bibitem{dolan}
B.~P.~Dolan, Class.\ Quantum\ Grav. {\bf 28}, 235017 (2011). 

\bibitem{haykoda}
H.~Kodama, Prog.\ Theor.\ Phys. {\bf 63}, 1217 (1980); S.~A.~Hayward, Class.\ Quantum\ Grav. {\bf 15}, 3147 (1998); Phys.\ Rev.\ D {\bf 49}, 6467 (1994); Phys.\ Rev.\ D {\bf 53}, 1938 (1996). 

\bibitem{jacobson}
T.~Jacobson, Phys.\ Rev.\ Lett. {\bf 75}, 1260 (1995).

\bibitem{helo0}
P.~Bin\'etruy and A.~Helou, Class.\ Quantum\ Grav. {\bf 32}, 205006 (2015).

\bibitem{bekenstein2}
J.~D.~Bekenstein, Phys.\ Rev.\ D {\bf 23}, 287 (1981); Int.\ J.\ Theor.\ Phys. {\bf 28}, 967 (1989).

\bibitem{helou}
A.~Helou, arXiv:1502.04235 [gr-qc]; arXiv:1505.07371 [gr-qc].

\bibitem{posit}
D.~Wenjie-Tian and I.~Booth, Phys.\ Rev.\ D {\bf 92}, 024001 (2015).

\bibitem{modified}
Shi-Bei~Kong, H.~Abdusattar, Y.~Yin, H.~Zhang and Ya-Peng~Hu, Eur.\ Phys.\ J.\ C {\bf 82}, 1047 (2022); H.~Abdusattar, arXiv:2304.08348 [gr-qc].

\bibitem{various}
Y.~Gong and A.~Wang, Phys.\ Rev.\ Lett. {\bf 99}, 211301 (2007); Rong-Gen~Cai and S.~P.~Kim, J.\ High\ Energy\ Phys. {\bf 0502}, 050 (2005); T.~Kobayashi, J.\ Cosmol.\ Astropart.\ Phys. \textbf{07}, 013 (2020).

\bibitem{callen}
H.~B.~Callen, {\it Thermodynamics and an introduction to Thermostatistics}, John Wiley, (1985).

\bibitem{colgain}
E.~O.~Colg\'ain and M.~M.~Sheikh-Jabbari,  Class.\ Quantum\ Grav. {\bf 38}, 177001 (2021).

\bibitem{hp}
S.~Wang, Y.~Wang and M.~Li, Phys.\ Reports {\bf 696}, 1 (2017).

\bibitem{Nojiri:2005pu}
S.~Nojiri and S.~D.~Odintsov,
Gen. Rel. Grav. \textbf{38}, 1285-1304 (2006).

\bibitem{Nojiri:2019yzg}
S.~Nojiri, S.~D.~Odintsov and E.~N.~Saridakis,
Nucl. Phys. B \textbf{949}, 114790 (2019).

\bibitem{Nojiri:2017opc}
S.~Nojiri and S.~D.~Odintsov,
Eur. Phys. J. C \textbf{77}, no.8, 528 (2017).

\bibitem{qft}
I.~L.~Shapiro and J.~Sola, Phys.\ Lett.\ B {\bf 682}, 105 (2009).

\bibitem{cohen}
A.~Cohen, D.~Kaplan and A.~Nelson, Phys.\ Rev.\ Lett. {\bf 82}, 4971 (1999).

\bibitem{Hsu:2004ri}
S.~D.~H.~Hsu,
Phys.\ Lett.\ B \textbf{594}, 13 (2004).

\bibitem{li}
M.~Li, Phys.\ Lett.\ B {\bf 603}, 1 (2004).

\bibitem{transient1}
M.~Cruz, S.~Lepe and G.~E.~Soto, Phys.\ Rev.\ D {\bf 106}, 103508 (2022).

\bibitem{transient2}
M.~Cruz and S.~Lepe, Phys.\ Dark\ Univ. {\bf 42}, 101367 (2023).

\bibitem{deHaro:2023lbq}
J.~de Haro, S.~Nojiri, S.~D.~Odintsov, V.~K.~Oikonomou and S.~Pan,
Phys. Rept. \textbf{1034}, 1-114 (2023).

\bibitem{Trivedi:2023zlf}
O.~Trivedi,
Symmetry \textbf{16}, no.3, 298 (2024).

\bibitem{Trivedi:2022ngt}
O.~Trivedi and M.~Khlopov,
Phys. Dark Univ. \textbf{36}, 101041 (2022).

\bibitem{cubic}
G.~Arciniega, J.~Edelstein and L.~G.~Jaime, Phys.\ Lett.\ B {\bf 802}, 135272 (2020); L.~G.~Jaime and G.~Arciniega, Phys.\ Lett.\ B {\bf 827}, 136939 (2022). 

\bibitem{sixth}
C.~Erices, E.~Papantonopoulos and E.~N.~Saridakis, Phys.\ Rev.\ D {\bf 99}, 123527 (2019).

\bibitem{oliva}
A.~Cisterna, N.~Grandi and J.~Oliva, Phys.\ Lett.\ B {\bf 805}, 135435 (2020).

\bibitem{h6}
H.~Abdusattar, Shi-Bei~Kong, H.~Zhang and Ya-Peng~Hu, Phys.\ Dark\ Univ. {\bf 42}, 101330 (2023).

\bibitem{vanderwaals}
D.~C.~Johnston, {\it Advances in Thermodynamics of the van der Waals Fluid}, Morgan \& Claypool Publishers, (2014).

\bibitem{diaz}
E.~N.~Saridakis, P.~F.~Gonz\'alez-D\'\i az and C.~L.~Sig\"uenza, Class.\ Quantum\ Grav. {\bf 26}, 165003 (2009).

\bibitem{grandaoliveros}
L.~N.~Granda and A.~Oliveros, Phys.\ Lett.\ B {\bf 669}, 275 (2008); Phys.\ Lett.\ B {\bf 671}, 199 (2009).

\bibitem{ricci}
Changjun~Gao, Fengquan~Wu, Xuelei~Chen and You-Gen~Shen, Phys.\ Rev.\ D {\bf 79}, 043511 (2009).

\bibitem{ricci2}
L.~P.~Chimento, M.~Forte, M.~G.~Richarte, Mod.\ Phys.\ Lett.\ A {\bf 28}, 1250235 (2013).

\bibitem{qcd}
F.~R.~Urban and A.~R.~Zhitnitsky, Phys.\ Lett.\ B {\bf 688}, 9 (2010); N.~Ohta, Phys.\ Lett.\ B {\bf 695}, 41 (2011); A.~R.~Zhitnitsky, Phys.\ Rev.\ D {\bf 92}, 043512 (2015); A.~O.~Barvinsky and A.~R.~Zhitnitsky, Phys.\ Rev.\ D {\bf 98}, 045008 (2018); B.~Holdom, Phys.\ Lett.\ B {\bf 697}, 351 (2011); A.~Yamamoto, Phys.\ Rev.\ D {\bf 90}, 054510 (2014); Rong-Gen~Cai, Zhong-Liang~Tuo, Hong-Bo~Zhang and Qiping~Su, Phys.\ Rev.\ D {\bf 84}, 123501 (2011); Rong-Gen~Cai, Zhong-Liang~Tuo, Ya-Bo~Wu and Yue-Yue~Zhao, Phys.\ Rev.\ D {\bf 86}, 023511 (2012); M.~Biswas, U.~Debnath, S.~Ghosh and B.~K.~Guha, Eur.\ Phys.\ J.\ C {\bf 79}, 659 (2019); M.~Rezaei, J.~Solà-Peracaula and M.~Malekjani, Mon.\ Not.\ Roy.\ Astron.\ Soc. {\bf 509}, 2593 (2021); H.~Hossienkhani, H.~Yousefi, N.~Azimi and Z.~Zarei, Astrophys.\ Space\ Sci. {\bf 365}, 59 (2020).

\bibitem{hdemodels}
E.~N.~Saridakis, Phys.\ Rev.\ D {\bf 102}, 123525 (2020); M.~Tavayef, A.~Sheykhi, K.~Bamba and H.~Moradpour, Phys.\ Lett.\ B {\bf 781}, 195 (2018); H.~Moradpour, S.~A.~Moosavi, I.~P.~Lobo, J. P. Morais-Graça, A.~Jawad and I.~G.~Salako, Eur.\ Phys.\ J.\ C {\bf 78}, 829 (2018); A.~Sayahian~Jahromi, S.~A.~Moosavi, H.~Moradpour, J.~P.~Morais-Graça, I.~P.~Lobo, I.~G.~Salako and A.~Jawad, Phys.\ Lett.\ B {\bf 780}, 21 (2018).

\bibitem{stability}
Y.~S.~Myung, Phys.\ Lett.\ B {\bf 652}, 223 (2007).

\bibitem{unstable}
A.~Oliveros, M.~A.~Sabogal and Mario~A.~Acero, Eur.\ Phys.\ J.\ Plus {\bf 137}, 783 (2022).
\end{thebibliography}
\end{document}